\documentclass[12pt,preprint]{aastex}
\usepackage[numberedappendix]{emulateapj5}
\shorttitle{HST/WFPC\,2 INVESTIGATION OF NGC\,891}
\shortauthors{ROSSA ET AL.}
\begin{document}

\submitted{THE ASTRONOMICAL JOURNAL, in press (August 2004 issue)}
\title{A HUBBLE SPACE TELESCOPE WFPC\,2 INVESTIGATION OF THE DISK-HALO 
INTERFACE IN NGC\,891\altaffilmark{1}}

\author{J\"orn Rossa\altaffilmark{2} and Ralf-J\"urgen Dettmar}
\affil{Astronomisches Institut, Ruhr-Universit\"at Bochum, 
Universit\"atsstrasse 150/NA7,\\
D-44780 Bochum, Germany}
\email{jrossa@stsci.edu and dettmar@astro.rub.de}

\author{Ren\'e A. M. Walterbos}
\affil{New Mexico State University, Department of Astronomy, MSC\,4500, 
Box\,30001,\\
Las Cruces, NM 88003}
\email{rwalterb@nmsu.edu}

\and

\author{Colin A. Norman\altaffilmark{2}}
\affil{Department of Physics and Astronomy, Johns Hopkins University, 
3400 North Charles Street,\\ 
Baltimore, MD 21218}
\email{norman@stsci.edu}

\altaffiltext{1}{Based on observations with the NASA/ESA Hubble Space 
Telescope, obtained at the Space Telescope Science Institute, which is 
operated by the Association of Universities for Research in Astronomy, Inc. 
under NASA contract No. NAS5-26555. These observations are associated with 
proposal \#6588.}
%\altaffiltext{2}{Present address: Space Telescope Science Institute, 3700 
%San Martin Drive, Baltimore, MD 21218}
\altaffiltext{2}{Space Telescope Science Institute, 3700 San Martin Drive, 
Baltimore, MD 21218}

\begin{abstract}

We present deep narrowband observations with high spatial resolution of 
extraplanar diffuse ionized gas in the halo of NGC\,891, obtained 
with the WFPC2 on-board the HST. Our H$\alpha$ observations, centered on 
the northern part of NGC\,891, reveal an extended gaseous halo, which fills 
almost the entire field of view of our WFPC2 observation. Whereas NGC\,891 
has been studied extensively with ground-based telescopes, here the small 
scale structure of the extended emission line gas is presented at high 
spatial resolution of 0\farcs1, corresponding to 4.6\,pc at the distance to 
NGC\,891. The majority of the H$\alpha$ emission is diffuse. Several 
discrete features (e.g.,\,filaments) are visible as well. Some of the 
filaments reach distances of up to 2.2\,kpc above the galactic plane, and are 
extremely collimated, even at high galactic latitudes. We compare the 
morphology of these filaments with theoretical models, which describe 
possible transport mechanisms in a general way. Despite the prominent dust 
lane, many bubbles, shells and super-shells can be discerned in the midplane. 
We also investigate extraplanar dust features, which are best visible in 
unsharp-masked images of our broadband F675W image, and we compare them to 
the spatial distribution of DIG filaments. The high-$|z|$ dust is detected 
out to distances of 2.2\,kpc above/below the galactic midplane. Individual 
dust features, however, are not spatially correlated with diffuse ionized gas 
counterparts, such as individual filaments. Quite interestingly, the 
orientation of the dust features changes from being mostly aligned 
perpendicular to the disk at low galactic latitudes, to a parallel alignment 
at high $|z|$. We compare the diffuse ionized gas distribution to the hot 
ionized medium, traced by X-ray observations performed by Chandra. There 
exists a good correlation of the presence of the warm and hot ionized gas, in 
particular, an X-ray bright region at $|z|\sim$1-1.5\,kpc fills the entire 
northern halo region, whereas the intensity in the midplane is considerably 
depressed. We also compare the sizes of individual H$\alpha$ emission line 
features in the midplane of NGC\,891 with similar structures that are known 
in our Milky Way and in the LMC.

\end{abstract}


\section{INTRODUCTION}
\label{s:intro}

The detection of extended diffuse ionized gas (DIG) or frequently also called 
warm ionized medium (WIM) in halos of late-type spiral galaxies is generally 
believed to be correlated with the star formation activity in the galaxy 
disk. This $\rm{H}^+$ layer in external galaxies is equivalent to the 
{\em Reynolds layer}, which is observed in our Milky Way \citep{Rey84}. More 
recent galactic studies showed that the diffuse H$\alpha$ emission is 
ubiquitous and detected in almost every direction on the sky \citep{Haf03}.
DIG has a typical electron density of $n_{\rm e} \approx 0.1\,\rm{cm^{-3}}$ 
in the disk (decreasing exponentially towards higher galactic latitudes), and 
has a typical scale height of 1-2\,kpc. 

In the theoretical picture, the gas, most likely driven by collective 
supernovae, is expelled into the halos of these galaxies. \citet{NoIk} 
developed the theory in which the gas is being transported through 
tunnel-like features into the halo, called {\em chimneys}, provided the 
superbubbles created from SNe fulfill the breakout criterion. Depending on 
the strength of the gravitational potential the gas may be able to fall back 
onto the galactic disk, which is described as the {\em galactic fountain} 
scenario \citep{Sha76,Avi00}. A few other investigations of the multiphase 
interstellar medium involving SNe have been conducted 
\citep[e.g.,\,][]{Ko99,WaNo99,WaNo01,Wa01}.

The fractional contribution of DIG to the total H$\alpha$ emission in 
galaxies reaches typically values of $\approx$25-50\% \citep[e.g.,\,][]{
Wal94,Fer96b,LeHe95,Th02}, and DIG is thus an important constituent of the 
ISM. The emission is most likely due to photoionization by hot OB stars 
\citep{MiCo93,DoSh}. The temperatures, determined from upper limits of 
diagnostic line ratios (e.g.,\,[\ion{N}{2}] lines) in the galaxy and in a few 
external galaxies are in the range of 8000-13000\,K. \citep{Ra97,Rey01,Tu00}. 
However, other ionizing and heating mechanisms such as shock-ionization 
\citep{Ch85}, magnetic reconnection \citep{Bir98} and turbulent mixing layers 
\citep{Sl93} have been invoked in order to account for the observed emission 
line ratios, which are not consistent with current pure photoionization 
models \citep{Ma86,DoMa94}. The line ratios in the disk are more or less 
consistent with the models. However, recently determined line ratios in 
external galaxies, such as the [\ion{O}{3}]/H$\alpha$, increase with 
increasing galactic distance, which are not compatible with these models, as 
observations have shown \citep{Ra98,Ra00,Tu00}. 

Furthermore, it is presently not clear how the ionizing photons can travel 
such vast distances from the disk into the halo, if OB stars are regarded as 
the primary source. Gaseous halos seem to be a common feature among starburst 
galaxies \citep{LeHe95}, where the gas is most likely expelled into the halos 
by starburst- or SNe-driven superwinds. However, in normal or {\em 
quiescent} galaxies the presence of extended gaseous halos does not seem to 
be the general case but is mostly encountered in late type spirals with high 
SF activity on both local and global scales \citep{RD02a,RD02b}. 

A few investigations have been undertaken to study the gaseous halos in 
{\em quiescent} galaxies \citep{Pi94,Ra96,Ho99,Co00,RD2000,MiVe03}, but not 
all of these galaxies revealed extraplanar DIG (eDIG). The first larger 
survey to study eDIG in the disk-halo interaction (DHI) context has been 
carried out recently, which covered 74 edge-on galaxies \citep{RD02a,RD02b}. 
It was shown in this investigation that nearly 41\,\% of the survey galaxies 
have eDIG detections, and a minimal threshold of star formation rate per 
unit area was derived \citep{RD02a}. 

A variety of morphological features are generally observed (e.g.,\,filaments, 
plumes, bubbles, shells), and sometimes, in the case of actively star forming 
galaxies (e.g.,\,NGC\,4700), even a pervasive DIG layer at typical distances 
of 1-2\,kpc above/below the galactic plane was detected. Individual 
filaments, occasionally reaching distances of up to 6\,kpc, are superimposed 
in a few cases as well \citep{RD02b}. Whereas the above mentioned survey and 
the smaller samples studied by individual researchers have aimed at the 
detection and overall morphology of gaseous halos in edge-on galaxies, 
little is actually known about the small scale structure of the eDIG. This 
small scale structure can only be studied in the nearest galaxies with the 
highest spatial resolution available. Therefore, high resolution observations 
with the HST in nearby edge-on spirals, yielding spatial resolutions of the 
order of a few parsecs, such as in the case of NGC\,891, seem promising to 
study the disk-halo connection in detail.  

\begin{table*}
\begin{center}
\caption{Journal of observations\label{t:exposure}}
\begin{tabular}{cccccccc}
\tableline
\tableline
Object & R.\,A. (J\,2000.0) & Dec. (J\,2000.0) & ${\rm{t_{exp}}}$ & Filter & 
$\lambda_{\rm{m}}$ & FWHM & p.a. \\
& [hh mm ss] & [\degr\,\,\arcmin\,\,\arcsec] & [s] & & [{\AA}] & [{\AA}] & 
[\degr] \\
\tableline
NGC\,891 & 02 22 39 & +42 22 11 & $4\times$400 & F675W & 6734.5 & 889.4 & 
22 \\
NGC\,891 & 02 22 39 & +42 22 11 & $8\times$2800 & F656N & 6561.5 & 22.0 & 
22 \\
NGC\,891 & 02 22 39 & +42 22 11 & $1\times$2700 & F656N & 6561.5 & 22.0 & 
22 \\
\tableline
\end{tabular} 
\end{center}
\end{table*}

\section{NGC\,891 AND THE DISK-HALO CONNECTION}
\label{s:ngc891}
   
The famous nearby edge-on spiral NGC\,891 is considered to be a twin to our 
own Milky Way in many respects. It is one of the best studied edge-on 
galaxies, which has been investigated in almost all wavelength regimes from 
the radio \citep[e.g.,\,][]{Da94} to the X-ray \citep{Br94} regime in the 
DHI context. About a decade ago an eDIG layer has been discovered in this 
galaxy (similar to the Reynolds layer in our Milky Way) by \citet{De90} and 
\citet{Ra90}, making use of H$\alpha$ narrowband imaging. Further 
spectroscopic investigations confirmed and revealed diffuse emission up to 
5\,kpc above the galactic plane \citep{Ke91,De92,Ra97,Ra98}. From the imaging 
observations it was shown that the star formation activity in NGC\,891 is not 
equally distributed along the disk, since an asymmetrical distribution of the 
DIG layer is observed. The dominating SF activity is located in the northern 
part of NGC\,891, whereas in the southern part there is much less activity. 
This is consistent with radio continuum observations \citep{Da94}, which are 
not affected by dust. Therefore, extinction can be excluded as a cause of 
the depression of the eDIG layer in the southern part. Instead, it is rather 
a real consequence of the decreasing strength of SF activity within the disk 
of NGC\,891. Those inhomogeneities are not unusual among actively star 
forming galaxies. 

\begin{figure*}[th]
\centering
\includegraphics[width=2in,angle=270]{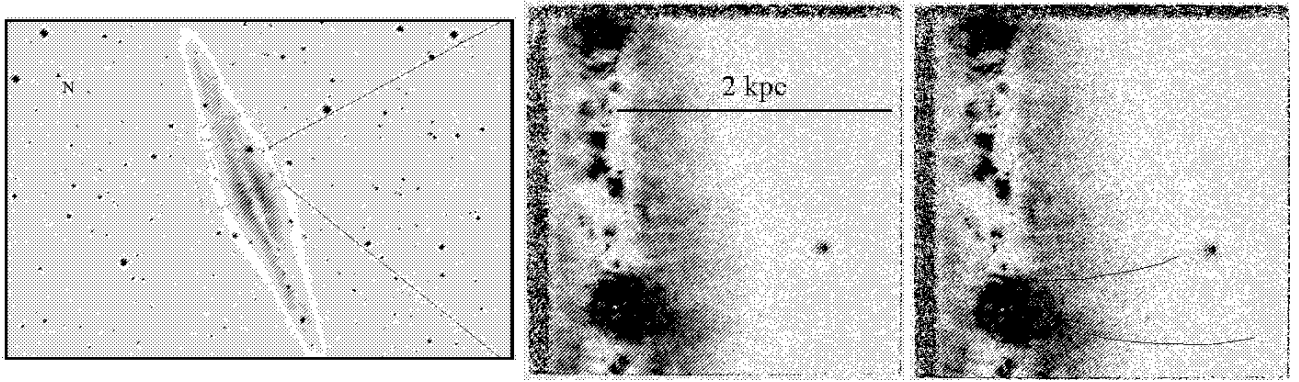}
\caption{NGC\,891: The left panel shows a DSS image with the WFPC2 field 
of view overlaid. The middle panel shows the drizzled continuum subtracted 
H$\alpha$ image of the disk-halo region just north of the bulge (WF3). There 
are individual long and thin filaments that emanate from a supergiant shell 
region in the galactic midplane. These filaments reach extraplanar 
distances of $\sim 2$ kpc. The filaments are clearly visible when viewed at 
different contrast on a computer screen, but might be difficult to view in 
the paper version. For clarity, the right panel therefore shows the same 
image as the middle panel, but with the filaments overdrawn in black for 
easy visibility. The most prominent filaments are very narrow with sizes 
of $\approx$15\,pc. Such morphological structures are not predicted by 
current theoretical models. Note, that the images in the middle and right 
panel are rotated by 22$\degr$, so north is to the NW 
quadrant.\label{f:gallery}}
\end{figure*}

In addition of bearing a prominent eDIG layer, NGC\,891 also shows 
spectacular extraplanar dust features, which reach up to $z\approx2$\,kpc 
into the halo as well \citep{HoSa97}. Correlation studies of the DIG 
distribution, traced by the H$\alpha$ emission, so far do exist with the 
radio continuum \citep{Da94}, and \ion{H}{1} \citep{Sw97}, and for the hot 
ionized medium (HIM), as observed in the X-rays \citep{Br94}. A previous 
study by \citet{HoSa99} showed that there is generally no direct correlation 
between the DIG filaments and extraplanar dust features on a small scale. 
This behavior was confirmed by \citet{RD02b} using a much larger sample. A 
re-investigation of the extraplanar dust in NGC\,891 using deeper broad 
band images also validates this trend \citep{HoSa00}. FIR and sub-mm 
observations, as performed with the ISO satellite and with SCUBA at the JCMT, 
respectively, have also detected amounts of extended cold dust in NGC\,891 
\citep{Alton98}. In addition, an extended halo has been reported from CO 
observations, obtained with the IRAM 30m telescope \citep{Ga92}. All these 
observations strongly imply that there is a large scale circulation of matter 
from the disk into the halo. This is directly linked to galaxy evolution, as 
the IGM is enriched with metals by those outflows. 

The distance to NGC\,891, as a member of the NGC\,1023 group, has been 
derived to be 9.5\,Mpc \citep{vdB92}, which is consistent with other distance 
determinations such as the Planetary Nebula Luminosity Function (PNLF), which 
yielded 9.9\,Mpc \citep{Ci91}. We will adopt a distance of 9.5\,Mpc 
throughout this paper. With this distance NGC\,891 is a prime candidate for 
high resolution studies of the small scale structure of the DIG. Furthermore, 
this galaxy has an inclination which is very close to be perfectly edge-on 
($i\gtrsim88\fdg6$), where the disk separates from the halo in projection 
very well. We therefore have carried out H$\alpha$ observations with the HST, 
making use of the WFPC2, which yielded a spatial resolution of 4.6\,pc at 
the distance of NGC\,891. This allowed us to study the distribution of DIG 
in external galaxies with unprecedented detail. Similar studies have been 
carried out recently for the starburst galaxy NGC\,3079 \citep{Ce01,Ce02}, 
where the nucleated gaseous outflowing cone was imaged and compared to the 
X-ray morphology, studied with Chandra. Furthermore, one additional 
HST/Chandra comparison study in the case of the interacting actively star 
forming Virgo spiral galaxy NGC\,4631 was performed \citep{Wan01}. 

\section{OBSERVATIONS AND OBSERVING STRATEGY}
\label{s:obs}

The HST observations of NGC\,891 were carried out with the WFPC2 camera. 
For the narrowband (H$\alpha$) observations the F656N filter was used, and 
the broadband images (to be used for continuum subtraction) were obtained 
with the F675W filter. The datasets have the following abbreviations 
U54Y0202B (narrowband) and U54Y0101B (broadband). In total, 8 HST orbits 
were performed to obtain a deep H$\alpha$ image of NGC\,891 resulting in an 
on-source integration time of 25.1\,ksec. The observational parameters are 
listed in Table~\ref{t:exposure}. 

The individual exposures were taken between February 17-19 1999 in dither 
mode, in order to approach more closely the diffraction limit of the 
telescope (since the WFPC2 CCD pixels are undersampled), and to allow good 
cosmic ray removal. The WFPC2 field was centered on R.\,A. 
${\rm{02^h22^m39^s}}$, and Dec. $+42\degr22'11''$, both J\,2000.0 at HST 
orientation angle $\phi = 67\degr$, centered on the PC frame. This resulted 
in a $p.a. = 22\degr$ on the sky, thus aligned along the major axis of 
NGC\,891. The field covers the northern part of NGC\,891, slightly north off 
the central region, including parts of the prominent bulge (see 
Figure~\ref{f:gallery}). The field position and p.a. were chosen for a 
variety of scientific reasons, which are briefly elucidated below.  

From the detections of the extraplanar gas layer in NGC\,891 in the early 
nineties it was known that there is an asymmetrical eDIG distribution, with 
the prominent gaseous halo in the northern part of NGC\,891 \citep[cf.\,][]
{De90,Ra90}. Due to the velocity field of NGC\,891 which was studied in 
\ion{H}{1} \citep{Ru91}, the northern part was also the most suitable choice 
for our investigation, as it covered the H$\alpha$ emission in the available 
F656N filter passband. The southern part of NGC\,891 would not be covered by 
the FWHM of the F656N filter. Furthermore, interesting features (filaments, 
\ion{H}{2} regions) are visible in ground-based images in this specific 
northern part, which can be studied now with higher angular resolution. 

A few orientations of the WFPC2 aperture onto NGC\,891 were possible in 
principle. However, due to the time-critical observations (NGC\,891 could 
only be observed in small periods of time twice a year), and a possible 
further delay of the WFPC2 observations during 1998 (due to problems 
associated with the coolant of the NICMOS camera), we were prompted to change 
our original proposed orientation. After analyzing possible solutions, taking 
into account the above mentioned constraints, three viable orientations, 
which would cover most of the desired areas in NGC\,891, remained. We finally 
decided to choose the $\phi = 67\degr$ (p.a. = 22\degr) orientation, since 
the other two orientations suffered from relatively bright stars near the 
edges of the WFPC2 aperture, which would lead to reflexions and 
ghost-images in the WFPC2 field. In order to avoid contamination of bright 
stars on the border of these regions ($\rightarrow$ dragon's breath), which 
would ultimately lead to difficulties in removing artifacts that might be 
caused by these, we neglected the other two orientations.  

\section{DATA REDUCTION}
\label{s:data}

The data were pipeline processed in the usual manner, including bias level 
correction and flat-fielding. The further reduction was performed using 
the drizzle package within IRAF, developed for dithered images 
\citep[][]{FrHo02}. First we applied a coarse cosmic ray removal 
using the {\em precor} task to the images. Then the dithered images were 
cross-correlated with one image acting as the reference image, both in the 
on-band and off-band exposures. Subsequently, we determined the shifts 
using {\em shiftfind} and {\em avshift}, choosing the WF2 field as a weight 
parameter. The average shifts determined in each of the four WFPC2 CCD 
chips for each individual exposure had values ranging from -5 to 16 pixels. 
In order to check that the count rates in the sub-fields did not 
significantly deviate from one image to another, we compared the 
values, obtained from two selected fields, as determined through the image 
statistics. The images did not show any significant deviation. Now each of 
the four PC/WF sub-images were processed separately. We then generated 
cosmic ray masks. The shifted sub-images were drizzled with a scale factor of 
0.5 to an embedded output image having the dimension of 2048 $\times$ 2048 
pixels. After {\em drizzling} the images were median combined. Then we {\em 
blotted} (the process of reverse drizzling) the median image in order to have 
images which match the positions and the size of the input images. These 
blotted images can be used to identify and mask cosmic rays in the original 
images. Finally the blotted images were drizzled onto a finer grid. In order 
to recover some resolution from the undersampled images we thus shrunk the 
drop size. For more details on the drizzling method we refer to \citet{Go98}. 

Since from the drizzled $2048\times2048$ images no mosaics could be generated 
within IRAF, we have used the smaller $800\times800$ images to create a 
$1600\times1600$ mosaic, which was used for the flux calibration comparison, 
and identifying the bright filaments, whereas the higher resolved drizzled 
images have been used to analyze the small-scale morphology of the DIG.

\begin{figure*}[th]
\centering
\includegraphics[width=4.0in,angle=0]{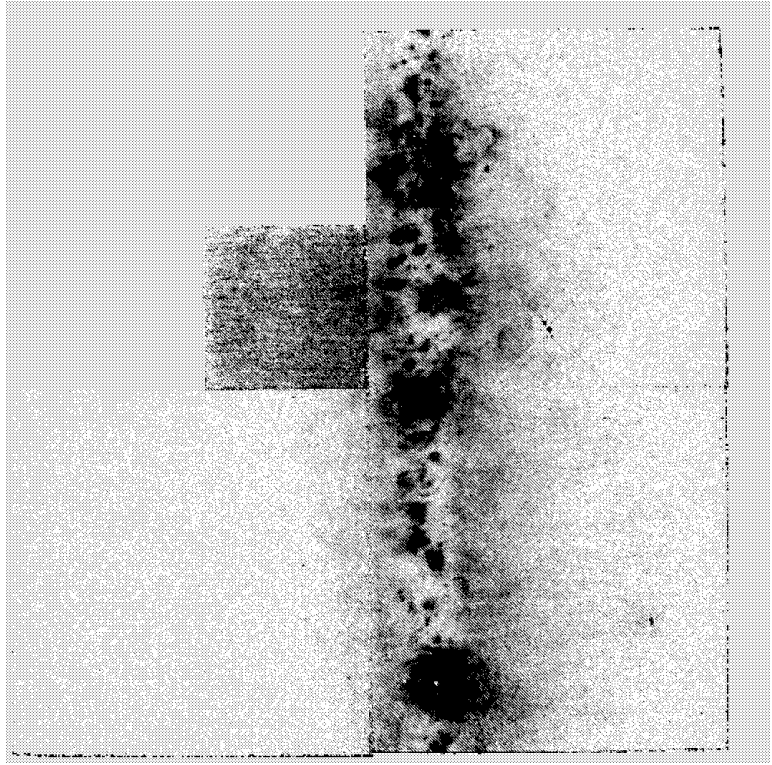}
\caption{Continuum subtracted F656N (H$\alpha$) image of NGC\,891, obtained 
after creating a mosaic from the four individual WF/PC fields. A very 
prominent DIG layer (z$\sim$1.8\,kpc) can be discerned, with individual 
superimposed filaments protruding from the disk region into the halo. The 
image size of this ``L''-shaped aperture measures $150''\times150''$, 
representing an area of $6.9\,\rm{kpc} \times 6.9\,\rm{kpc}$ at the distance 
of NGC\,891. The small PC-field measures $34''\times34''$. The spatial 
resolution is $0\farcs1$\,pix$^{-1}$.\label{f:hst891}}
\end{figure*}

\begin{figure*}[bh]
\centering
\includegraphics[width=4.0in,angle=0]{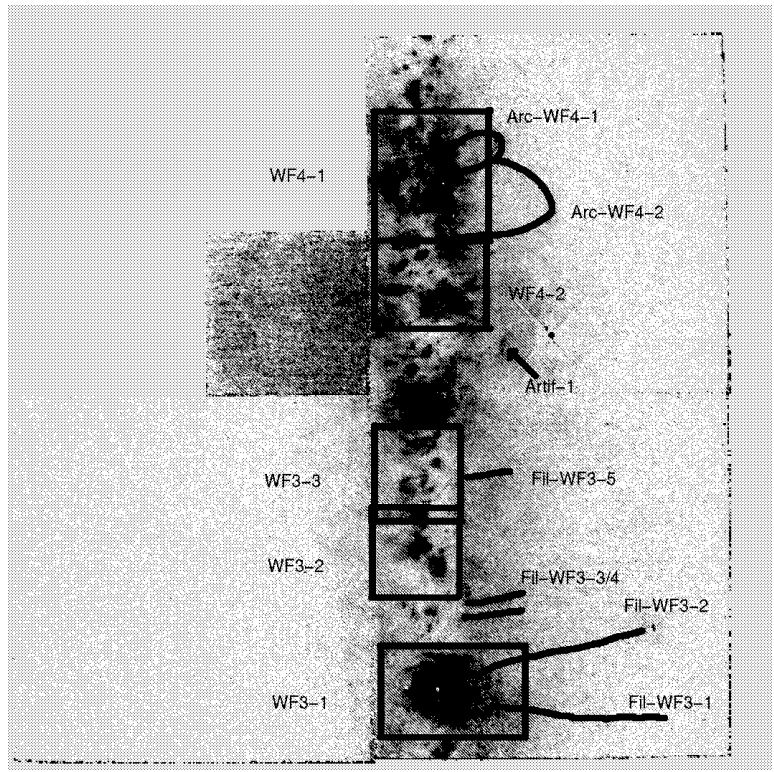}
\caption{Continuum subtracted F656N (H$\alpha$) image 
of NGC\,891, same as Figure~\ref{f:gallery}. Here several fields are marked, 
which are shown in the following figures in more detail. Furthermore, the 
most prominent filaments and arcs are indicated as well. The nomenclature is 
described in the text.\label{f:hst891d}}
\end{figure*}

\section{ANALYSIS}
\label{s:analysis}
\subsection{Narrowband imaging}
\label{ss:narrow}

Due to the given FWHM of 22.0\,{\AA} of the F656N filter, the filter passband 
also contains some amount of continuum. In order to obtain pure H$\alpha$ 
emission the F656N images had to be continuum subtracted by the scaled 
broadband filter (F675W) exposure. This was done by measuring the count 
rates of several individual stars, which were visible in all of the four 
WFPC2 fields in both the on-band and off-band exposures. The individual 
measurements were then averaged to obtain a final scaling factor. The 
off-band exposure, multiplied by the scaling factor, was subtracted from the 
on-band exposure to obtain a continuum free H$\alpha$ image. Due to the 
relatively small FWHM of the F656N filter there is no contamination of 
emission from the adjacent [\ion{N}{2}] doublet. Finally, a flux calibration 
has been performed. The net H$\alpha$ image's calibration was derived 
following \citet{Od99}. We did not include the correction they derive for 
contamination of [\ion{N}{2}] emission, since the velocity shift of NGC\,891 
places only little [\ion{N}{2}] emission in the filter passband. The 
calibration was consistent within the uncertainties with ground-based 
calibration \citep{Ra90}. The determined sensitivity of is of the order 
of $\rm{EM=20\,cm^{-6}\,pc}$ (rms per pixel). 

Generally speaking, using a broadband filter for the continuum subtraction 
is done to save exposure time in the offband filter. While the line emission 
is included in the broad filter passband, the contribution of line emission 
with respect to the continuum is very small. The equivalent widths of both 
filter passbands are quite different. Hence, when the continuum image is 
scaled to the narrowband image in order to bring the continuum levels 
together, the line emission in the broadband filter passband is scaled down 
quite substantially compared to the line emission in the narrowband filter, 
and very little (a few percent at most) of the line emission maybe removed 
from the narrowband filter passband. Since this data set is primarily geared 
to the morphology of the emission, and not to the diffuse or absolute levels, 
we opted to use the broadband method to cut back on required exposure time 
and to maximize the S/N.

A galactic contamination of the observed H$\alpha$ emission in extragalactic 
sources is feasible in principle, specifically for targets in areas of the 
sky close to the galactic plane. For details on the galactic H$\alpha$ 
emission we refer to the emission line surveys such as WHAM \citep{Haf03}, 
the Virginia Tech Spectral-Line Survey (http://www.phys.vt.edu/$\sim$halpha/) 
and SHASSA \citep{Ga01}. Galactic emission could occur, in that 
[\ion{N}{2}] emission from the Milky Way falls in the HST filter passband. 
However, this is unlikely to be a problem. We are discussing a very small 
region of the sky, and the Galactic emission could only make NGC\,891 appear 
diffuse if it were structured exactly such that it appeared concentrated 
towards NGC\,891's disk, which is what we see in the HST and ground-based 
data. That would seem to be a rather unlikely situation. However, we like to 
mention that there is at least one reported case (ESO209-9), where this can 
be an issue \citep{RD02b}. We note that ESO209-9 is closer to the Milky Way 
galactic plane in projection, and also the passband of the narrow filter used 
in that study was much larger than the current one of our WFPC2 study.

It should be noted that we cannot make a determination of the total extend 
of the NGC\,891 emission due to the relatively small field of view covered 
by the WFPC2 observation. For the same reason, we cannot determine the zero 
level background from our data. This can be better addressed by ground-based 
imaging and long-slit investigations \citep[e.g.,][]{De90,Ra90,Ra97,Ho99}.

\begin{table*}
\begin{center}
\caption{Properties of individual eDIG features identified in the continuum
free H$\alpha$ images\label{t:edigfil}}
\begin{tabular}{lccccc}
\tableline
\tableline
Identifier & $\Delta$\,x & $\Delta$\,y & $|z|$ & Dimensions & 
Classification\\ 
Fil-WFi-j & [$''$] & [$''$] & [pc] & [pc $\times$ pc] & \\
\tableline
Fil-WF3-1 & $-42.1$ & $-84.3$
& 2197$\pm$5 & $1561\pm5\times15\pm5$ & filament \\
Fil-WF3-2 & $-44.4$ & $-75.7$
& 1967$\pm$5 & $1552\pm5\times46\pm5$ & filament \\
Fil-WF3-3 & $-39.1$ & $-68.5$
& $~\,921\pm$5 & $~\,410\pm5\times46\pm5$ & filament \\
Fil-WF3-4 & $-38.2$ & $-68.7$
& $~\,930\pm$5 & $~\,502\pm5\times78\pm5$ & filament \\
Fil-WF3-5 & $-36.2$ & $-34.3$
& $~\,760\pm$5 & $~\,451\pm5\times46\pm5$ & filament \\
WF4-EH2 & $-42.4$ & $+29.1$
& $~\,617\pm$5 & $~\,~\,64\pm5\times55\pm5$ & extraplanar
\ion{H}{2} region \\
Arc-WF4-1 & $-45.2$ & $+35.9$
& $~\,760\pm$5 & $~\,460\pm5\times98\pm5$ & arc  \\
Arc-WF4-2 & $-56.5$ & $+20.4$
& 1271$\pm$5 & $~\,815\pm5\times55\pm5$ & arc \\
Fil-WF3-6 & $-20.2$ & $-63.7$        
& $~\,755\pm$5 & $~\,373\pm5\times92\pm5$ & filament \\
Fil-WF3-7 & $-18.8$ & $-54.0$
& 1368$\pm$5 & $~\,921\pm5\times60\pm5$ & filament \\
Fil-WF3-8 & $-15.8$ & $-18.7$
& 1257$\pm$5 & $~\,713\pm5\times50\pm5$ & filament \\ 
\tableline
\end{tabular}                   
\end{center}
\end{table*}
  
\subsection{Broadband imaging}
\label{ss:broad}

The broadband (F675W) images of NGC\,891 were obtained for two purposes. 
The primary aim was to subtract the continuum emission from the on-band 
images. They also allow a comparison of extraplanar dust features with eDIG 
filaments. These two physically distinct phases \citep[see\,][]{HoSa00} of 
the ISM can be studied in detail, to search for possible correlations between 
the two extraplanar ISM constituents. To better enhance the contrast between 
the dust structures and the background light from the galaxy, we have 
constructed unsharp-masked versions of our broadband images. These were 
obtained by dividing the broadband images by a smoothed version of the 
broadband images. The latter ones were created by using a Gaussian filtering 
technique. 
  
\section{RESULTS}
\label{s:results}
\subsection{Extraplanar diffuse ionized gas (eDIG)}
\label{ss:edig}

\begin{figure*}[th]
\centering
\includegraphics[width=4.0in,angle=270]{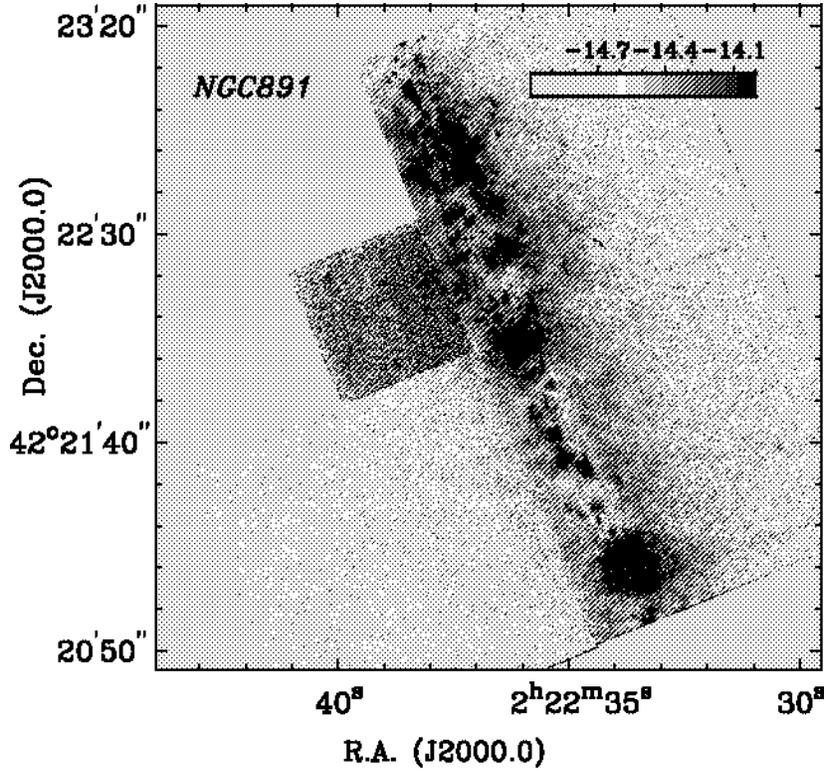}
\caption{Continuum subtracted F656N (H$\alpha$) image 
of NGC\,891. The calibrated image is shown with a coordinate grid and is 
plotted in a logarithmic stretch in units of 
$\rm{erg\,s^{-1}\,cm^{-2}\,pix^{-1}}$.\label{f:hst891log}}
\end{figure*}

The overall distribution of eDIG in the northern part of NGC\,891, as derived 
from our WFPC2 observations, shows a very smooth pattern (see 
Figure~\ref{f:hst891}). These narrowband HST observations offer the highest 
spatial resolution of eDIG studied in external {\em non-starburst} galaxies 
to date, with a spatial resolution of 4.6\,pc at the distance to NGC\,891. 
Whereas the majority of the high-$|z|$ H$\alpha$ emission is clearly diffuse, 
several filaments, however, can be discerned as well. A description of the 
various eDIG/DIG features (e.g.,\,filaments, plumes / bubbles, \ion{H}{2} 
regions, supershells) including the sizes, distances from the galactic 
midplane, are summarized in Table~\ref{t:edigfil} and Table~\ref{t:digplane}. 
The nomenclature of our identified eDIG filaments is in the following format: 
Fil-WF field + No., and selected fields. The brightest filaments are marked 
in Figure~\ref{f:hst891d}. The positional offsets $\Delta$\,x and $\Delta$\,y 
(in arcseconds), relative to the center of the PC field, are listed as well. 
The coordinates for the filaments refer to the observed position from which 
they seem to originate. For the two arcs the positional offsets refer to the 
outermost extension of each arc. As the arcs are irregular in shape, we list 
the distance to the outermost detected emission, rather than defining a 
radius, which would be difficult to assess, due to their observed morphology.
In Figure~\ref{f:hst891log} we show the calibrated image in a logarithmic 
stretch and with a coordinate grid.

\subsection{Morphology of eDIG}
\label{ss:lscale}

\begin{figure*}[th]
\centering
\includegraphics[width=4in,angle=270]{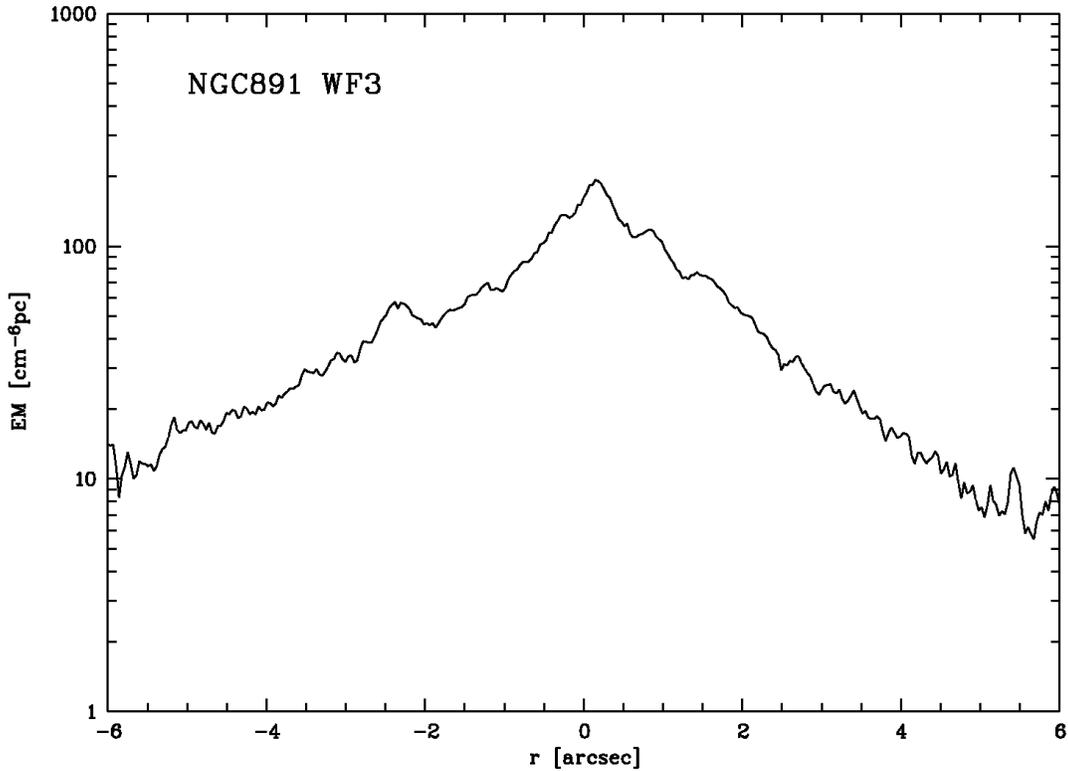}
\caption{Vertical profile of the H$\alpha$ emission in NGC\,891 (plotted 
logarithmically in units of EM = [$\rm{cm^{-6}\,pc}$]) perpendicular to the 
disk. The cut shows an averaged region (150\,pc wide), centered at the bright 
emission complex at the center of the WFPC2 image (slightly north of the 
WF3-3 box in Figure~\ref{f:hst891d}).\label{f:profile}}
\end{figure*}

On the larger scale the H$\alpha$ emission is observed mostly in diffuse 
emission. The eDIG can be detected out to $|z|\sim1.8$\,kpc above the 
galactic midplane. We are actually limited by the field size of the WFPC2 
and sensitiviy. Given the position of NGC\,891 onto the WFPC2 chip, the 
field size is about 2.5-4.5\,kpc from the midplane to the border of the chip 
on the western/eastern side. The intensity of the DIG is brightest above the 
most active disk star forming regions (e.g.,\,supergiant shells). A few 
bright filaments and arcs are protruding from the regions, where supershells 
are located into the halo. The two giant arcs which emanate from the bright 
emission complex in the WF4 field (containing the supershell WF-4-SS1) out 
to $|z|\sim$1.3\,kpc, connect on either side with bright emission regions in 
the disk region. In several cases we detected smaller filaments which appear 
to be originating from regions deep within the disk, where no bright emission 
regions are detected. However, this may be an effect of local extinction, as 
we cannot determine at which depth the filaments are actually connected to 
the the disk. There can be bright supershells, embedded in regions of higher 
extinction or these regions can be located in deeper layers within NGC\,891's 
disk. Due to the extinction effects we cannot judge which of the both 
scenarios is true. However, the presence of filaments appearing at $|z|\geq 
500$\,pc clearly shows the decreasing influence of extinction at higher 
galactic latitudes. Scattered light from \ion{H}{2} regions contributes 
only a minor part of the observed DIG emission, as it has been shown from 
modeling, that the contribution at $|z|=600$\,pc falls to about 10\,\% 
\citep{Fe96}. A profile of the H$\alpha$ emission at the central part 
emission region, perpendicular to the disk of NGC\,891, is shown in 
Figure~\ref{f:profile}, where the intensity in EM ($\rm{cm^{-6}\,pc}$) is 
plotted as a function of the spatial distance from the midplane.     

Two bright arcs are visible in the WF4 field of the WFPC2, where the larger 
one reaches a distance of $|z|\sim1.2$\,kpc. In the WF3 field two super-thin 
filaments protrude from the disk, and north of them three prominent and 
thicker filaments are clearly recognizable. On the other side of the disk 
several fainter filaments emanate from the disk, which are not as straight 
as the brightest filaments already mentioned.  

In a general sense there is a good correlation of filaments with SF regions 
in the disk observed. Not only the very thin and long filaments, which seem 
to protrude from the supershells, have connections with SF regions in the 
disk. For instance, there are two bright arcs in the WF4 field (see 
Figure~\ref{f:hst891}), which are connected with and actually trace back to 
the disk, indicative of a galactic fountain scenario. Many of the smaller 
filaments also seem to originate from bright SF regions within the disk. 
However, it should be noted that extinction eludes us from making firm 
conclusions on a clear connection in several cases.  

\subsection{Small scale structure of eDIG}
\label{ss:sscale}

\begin{figure*}[th]
\centering
\includegraphics[width=6in,angle=0]{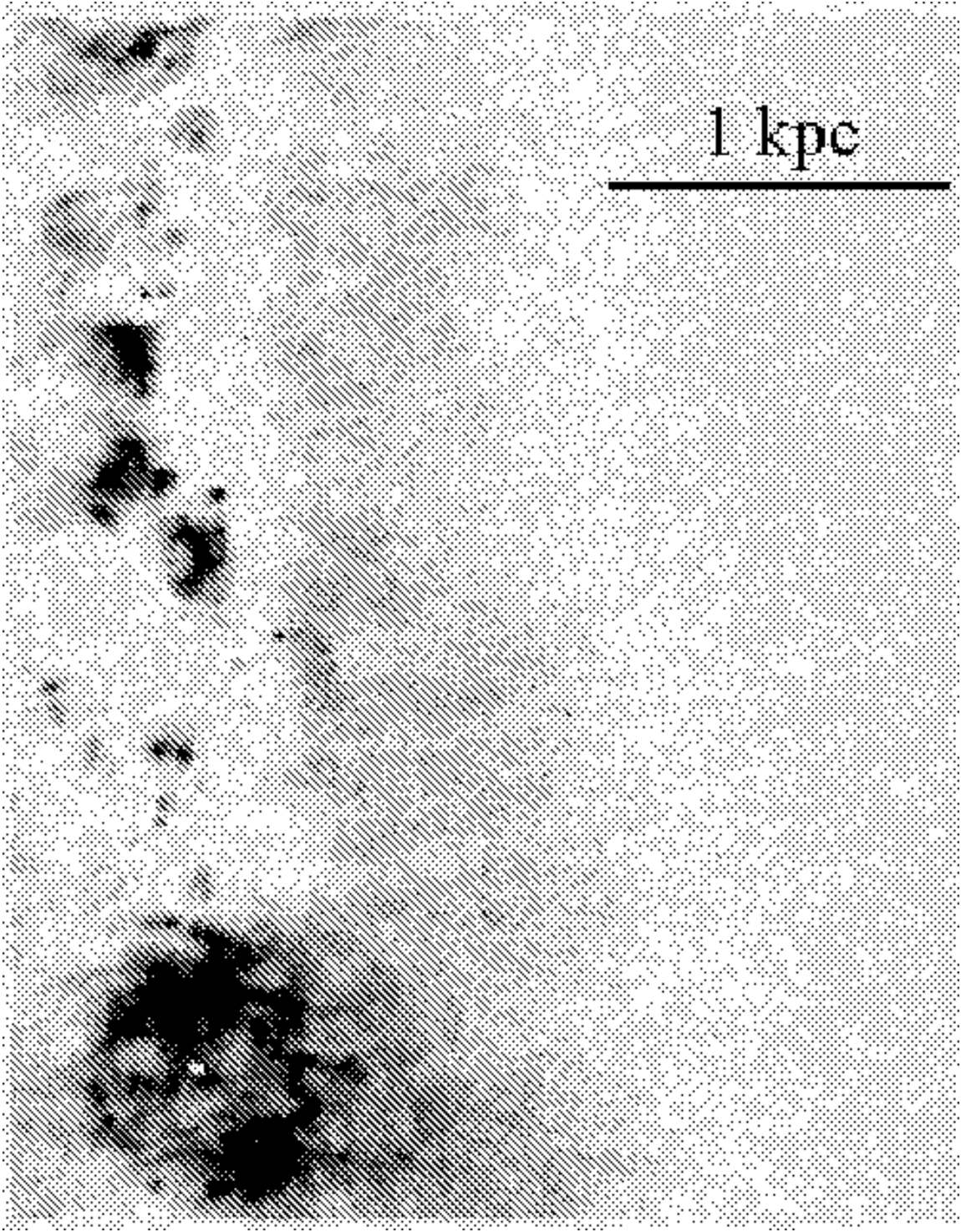}
\caption{Detailed view of the continuum subtracted H$\alpha$ DIG emission in 
the WF3 field. On the bottom the supergiant shell WF3-1-SS1 is seen along 
with other discrete features in the midplane. \label{f:enlarge1}}
\end{figure*}

\begin{figure*}[th]
\centering
\includegraphics[width=5.0in,angle=270]{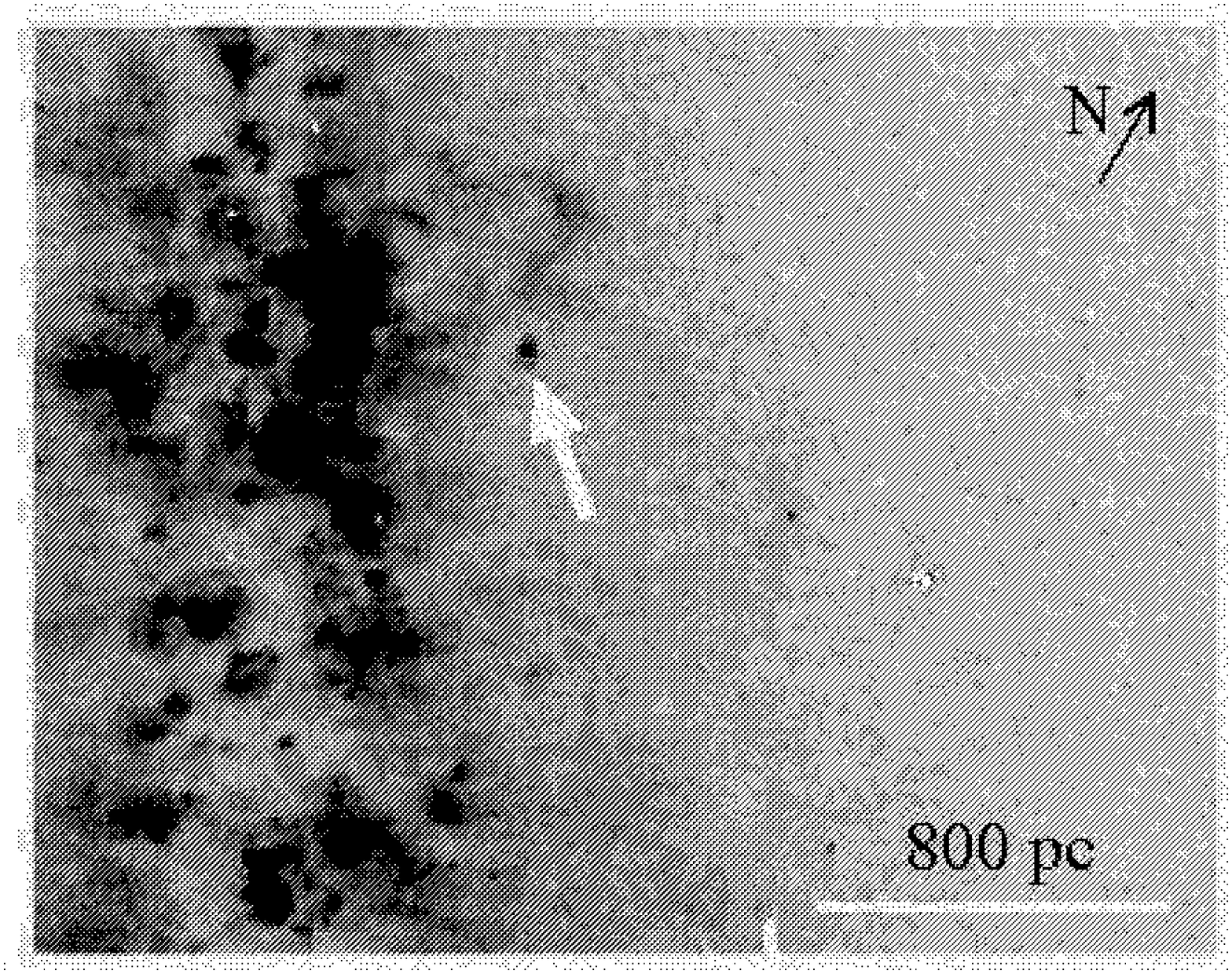}
%\plotfiddle{hst-wf4new.ps}{4.0in}{0}{1}{1}{200}{-200}
\caption{Detailed view of the continuum subtracted H$\alpha$ DIG emission in 
the WF4 field. Several filaments (some connecting back to the disk) can be 
discerned. The extraplanar \hbox{H\,{\sc ii}} region (WF4-EH2) is marked by 
the white arrow.\label{f:enlarge2}}
\end{figure*}

Although the eDIG distribution in NGC\,891 is mostly truly diffuse, numerous 
filaments and arcs are superimposed onto the extended DIG layer. The 
small scale structure of these filaments can be investigated down 
to sizes of about 5\,pc. There is no intricate filamentary network observed 
on the very smallest scales. From the sensitivities obtained we can conclude 
that a breakup of the diffuse emission can only occur below 
$\rm{EM=20\,cm^{-6}\,pc}$ (rms per pixel) into individual filaments. On the 
other hand, there are some surprisingly long and super-thin (collimated) 
filaments visible in the WF3 field (see Figure~\ref{f:gallery} and 
Figure~\ref{f:hst891}), which are slightly bent and very collimated even out 
to very high galactic latitudes. These filaments are barely resolved in 
ground-based images and have widths of 15\,pc to 50\,pc. The one filament 
(Fil-WF3-2) is coincident with the one previously identified by 
\citet{HoSa00} in ground-based WIYN images, which they have dubbed +026+035. 
From our HST observations we have estimated sizes that are a factor of two 
larger in length and a factor of three smaller in width. However, we note 
that our measurement refers to a mean width, as this filament does not have a 
constant width as a function of $|z|$. The curved feature marked in 
Figure~\ref{f:hst891d} as {\em Artif-1} on the WF4 chip is an artifact 
caused by the nearby star. In Figure~\ref{f:enlarge1} we show an enlargement 
of the WF4 field, where the two arcs can be seen in more detail. An 
extraplanar \ion{H}{2} region in the disk-halo interface (thick disk) at 
$|z|\sim620$\,pc is detected, which is indicated by an arrow in 
Figure~\ref{f:enlarge2} (see Section 6.6). 

\subsection{Structure of DIG features in the galactic midplane}
\label{ss:midplane}

Although NGC\,891 bears a rather prominent and extended dust lane which runs
across the galactic midplane (see Figure~\ref{f:dustall}), several emission 
regions which are located in areas not severely influenced by extinction 
(e.g.,\,individual shells, supergiant shells, \ion{H}{2} regions, and SNRs) 
are visible in our continuum subtracted H$\alpha$ images 
(Figure~\ref{f:enlarge1} and Figure~\ref{f:enlarge2}). The most spectacular 
feature in the galactic midplane is the supergiant shell WF3-1-SS1, which 
measures $677\,\rm{pc}\times824\,\rm{pc}$ (see Figure~\ref{f:enlarge1}). It 
is similar in size and shape as the supergiant shells found in the LMC 
\citep[e.g.,][]{Mea80}. A super-thin ($\Delta x \sim 15$\,pc) filament is 
protruding from this supergiant shell into the halo, and reaches a distance 
of $|z|\sim1.8$\,kpc. A listing of the most prominent emission features in 
the midplane is given in Table~\ref{t:digplane}, where information on the 
position within NGC\,891 and the physical sizes is given. Individual 
\ion{H}{2} regions are discerned (e.g.,\,Figure~\ref{f:gallery}), and a 
superbubble (see Figure~\ref{f:enlarge1}) can be identified. In 
Figure~\ref{f:enlarge1} and Figure~\ref{f:enlarge2} we have used different 
greyscale cuts to accentuate the inner structure more clearly.
          
In the continuum subtracted WFPC2 image (Figure~\ref{f:hst891}) there are 
local regions (i.e. between the two supershells) which show a less pronounced
SF activity, compared to most other regions. However, this could be 
attributed to either less SF, or the emission regions are embedded in deeper 
layers, and hence are more affected by extinction. Several filaments emerge 
from the disk at intermediate distances of about 400\,pc above the disk, and 
this might be suggestive that there are strong local inhomogeneities in the 
dust absorption. 

\begin{table*}
\begin{center}
\caption{Properties of emission features in the galactic midplane of 
NGC\,891\label{t:digplane}}
\begin{tabular}{lccccc}
\tableline
\tableline
Identifier & $\Delta$\,x & $\Delta$\,y & $|z|$ & Dimensions & Classification 
\\
WF-$i$+Ident. & [$''$] & [$''$] & [pc] & [pc $\times$ pc] & \\
\tableline
WF3-1-SS1 & $-31.9$ & $-79.1$ & 134$\pm$5 & $677\pm5\times824\pm5$ & 
supergiant shell \\
WF3-3-SS2 & $-27.8$ & $-19.8$ & ~\,55$\pm$5 & $700\pm5\times511\pm5$ & 
supergiant shell \\
WF3-SB-1 & $-34.1$ & $-31.8$ & 110$\pm$5 & $138\pm5\times124\pm5$ & 
superbubble \\
WF3-H2-1 & $-28.2$ & $-43.0$ & ~\,37$\pm$5 & $152\pm5\times203\pm5$ & 
\ion{H}{2} region \\
WF3-H2-2 & $-28.1$ & $-48.7$ & ~\,41$\pm$5 & $226\pm5\times230\pm5$ & 
\ion{H}{2} region \\
WF3-H2-3 & $-31.9$ & $-53.3$ & 134$\pm$5 & $166\pm5\times184\pm5$ & 
\ion{H}{2} region \\
WF4-SS-1 & $-32.7$ & $+30.8$ & 170$\pm$5 & $345\pm5\times244\pm5$ & 
supershell \\
\tableline
\end{tabular}                   
\end{center}
\end{table*}

\subsection{Extraplanar dust}
\label{ss:edust}

\begin{figure*}[th]
\centering
\vspace*{0.25in}
\includegraphics[width=7in,angle=0,clip=t]{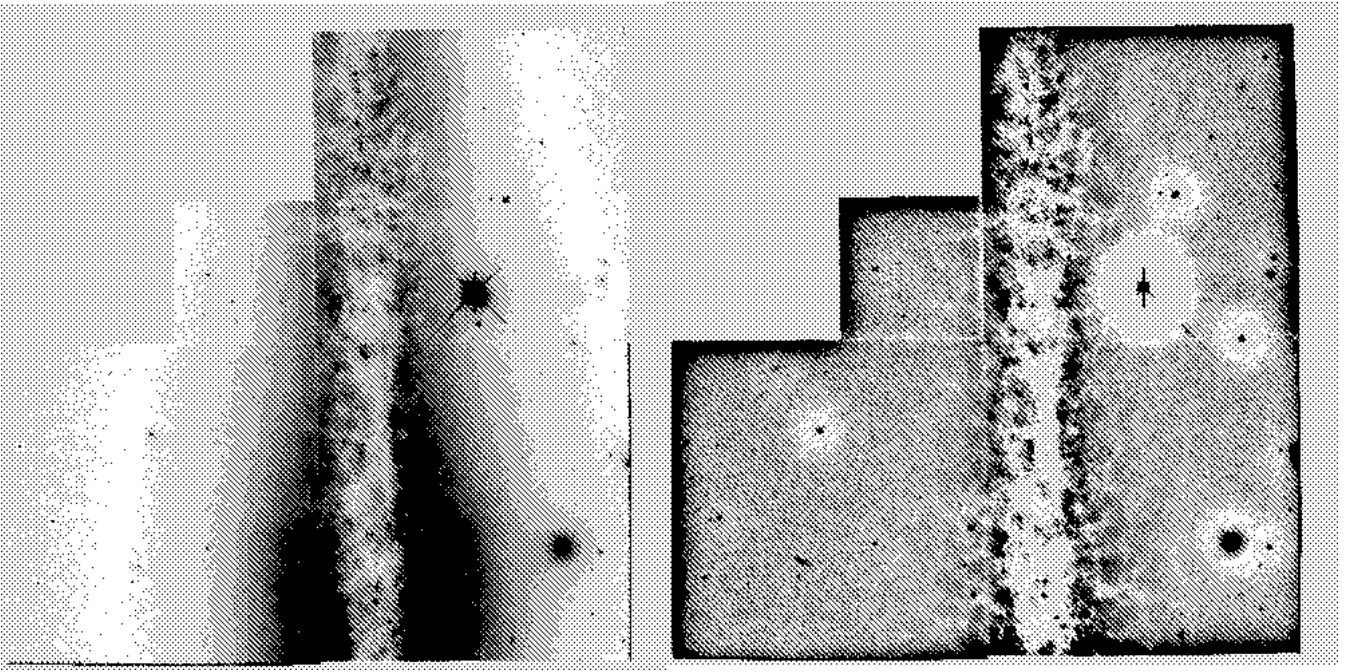}
\caption{The left panel shows the F675W image. The unsharp-masked broadband 
image of NGC\,891 is shown in the right panel, obtained after dividing the 
F675W image by a Gaussian-filtered F675W image. High-$|z|$ structures can be 
identified up to $\sim2$\,kpc. While the highest dust filaments seem to be 
aligned mostly parallel to the galactic disk, numerous dusty filaments in the 
region $|z|\sim 1-1.5$\,kpc are aligned perpendicular to the galactic 
disk.\label{f:dustall}}
\end{figure*}

\begin{figure*}[th]
\centering
\vspace*{0.15in}
\includegraphics[width=6.5in,angle=0]{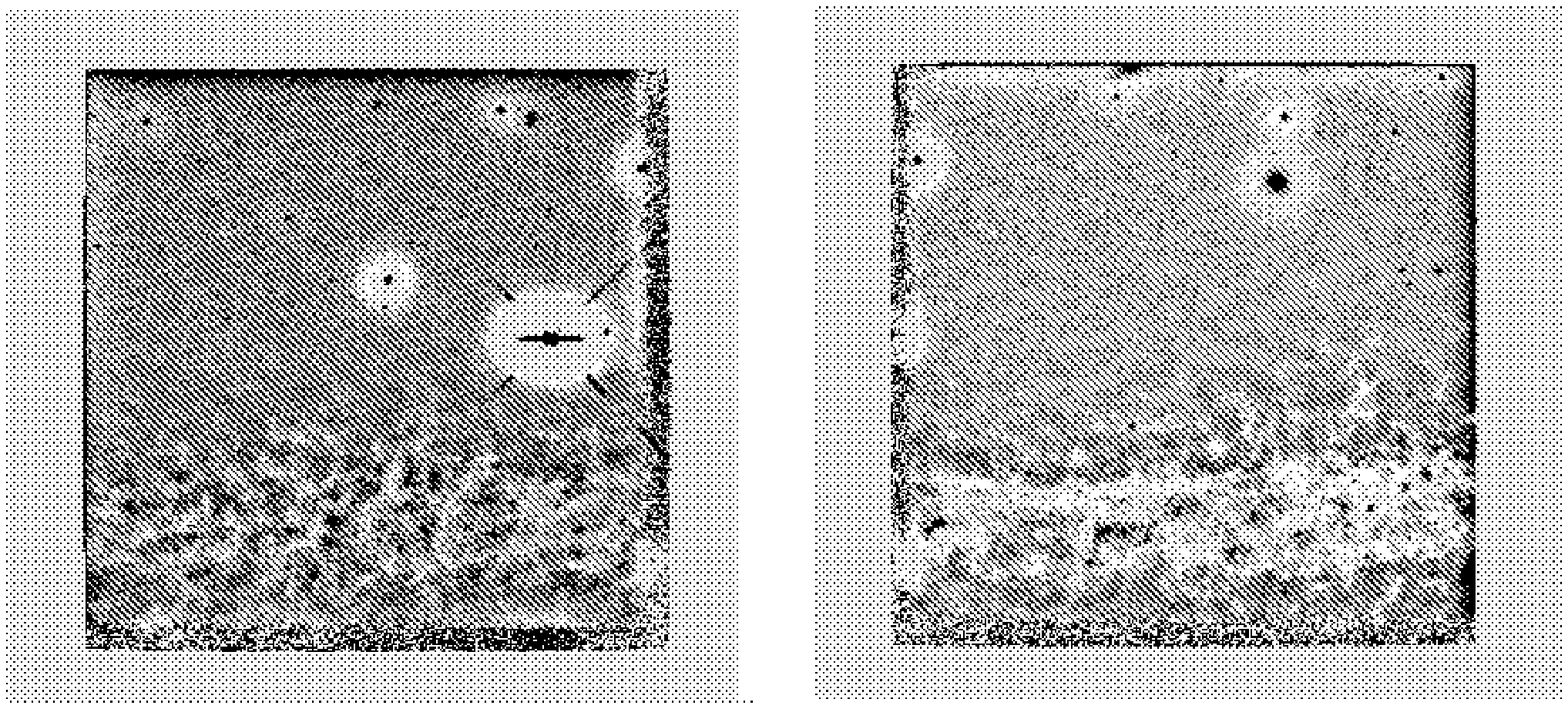}
\caption{Unsharp-masked F675W image of the WF4 field (left) and WF3 field 
(right), revealing many individual dust filaments. The image is oriented 
with the galactic plane being parallel (rotated by $90\degr$ to previous 
figures) for better convenience.\label{f:dustenl}}
\end{figure*}

The extraplanar dust in NGC\,891 reveals an intricate network of dusty 
filaments, which reach extraplanar distances of $|z| \sim 2$\,kpc. These 
dusty filaments and features are best visible in our unsharp-masked broadband 
image (Figure~\ref{f:dustall}). In Figure~\ref{f:dustenl} we show 
a more detailed view of the individual dust patches in the two WFPC2 
fields (WF3 + WF4), as the dust lane is covered in these two fields. As 
already studied with ground-based telescopes under exceptional good seeing 
conditions of $0\farcs6$, Howk \& Savage have mapped the extraplanar dust in 
NGC\,891 before in the B-band and V-band \citep{HoSa97,HoSa00} using the 
WIYN 3.5m telescope. We can confirm their results, and we detect basically 
the same structure for the extraplanar dust, although our observations were 
made in the R-band (F675W), which is somewhat more transparent to dust as 
their B- and V-band. However, the resolution of the WFPC2 is better by a 
factor of six, and therefore our dust filaments are better resolved. The 
structure of the smallest visible dust features on a scale of less than 
5\,pc can be resolved. The dust filaments show a pronounced sub-structure 
that is not well resolved in the WIYN images by \citet{HoSa00}. 
Interestingly, the high-$|z|$ dust shows basically two different alignments 
at various scale heights. The very extended high-$|z|$ dust at 
$|z|\sim2$\,kpc is aligned parallel to the galactic plane, while the dust 
features at intermediate distances $|z|\sim0.7-1.5$\,kpc are preferentially 
aligned perpendicular to the disk, although generally a superposition of 
various directions is observed. As \citet{HoSa97,HoSa00} have already given 
an extensive analysis of the individual extraplanar dust features in 
NGC\,891, we refrain here from discussing all further details again, as our 
main focus is on the DIG.

\subsection{Extraplanar \ion{H}{2} regions and halo PNe}
\label{ss:pn}

An interesting side aspect of the detection of H$\alpha$ emission in the 
disk-halo interface of spiral galaxies is the search for possible 
extraplanar \ion{H}{2} regions giving rise to star formation far above the 
galactic disk. Over the last couple of years a few detections of extraplanar 
and intracluster \ion{H}{2} regions in and surrounding spiral galaxies have 
been reported \citep[e.g.,][]{Fer96,Ge02,Tu03,Rya04}. Specifically, for 
NGC\,891 \citet{HoSa97} reported a detection of a \ion{H}{2} region in the 
disk-halo interface, which we also identify in our WFPC2 data, dubbed 
WF4-EH2. It is located at about $|z|\sim620$\,pc above the galactic plane 
(see Figure~\ref{f:enlarge2}). We identify a double morphology, with one knot 
being much brighter. The size we measured is approximately 45\,pc. We note 
that the knot is much smaller, with some extended diffuse emission around it. 
We derive a luminosity of $L=8\times10^{36}\,\rm{erg\,s^{-1}}$, which is 
slightly less than the luminosity of the Orion nebula \citep[e.g.,][]{Ke88}. 
There is a somewhat fainter blotch at a similar distance from the galactic 
midplane visible in that same figure just south of WF4-EH2, which is probably 
another \ion{H}{2} region. All these observations of extraplanar \ion{H}{2} 
regions establish a low level of star formation at high galactic latitudes. 
This should be considered in discussing the existence of early-type stars in 
the Milky Way halo \citep[e.g.,][]{Ki95,Ham96}. However, it should be 
mentioned that the occurrence of these \ion{H}{2} regions is a relatively 
rare phenomenon, hence they are unlikely to contribute much, if any, to 
the ionization of DIG above the disk. Considering that some or all of the 
\ion{H}{2} regions are due to runaway O-stars is very unlikely. A runaway 
O-star would no longer have an associated \ion{H}{2} region anymore.

\begin{figure*}[th]
\centering
\includegraphics[width=7in,angle=0,clip=t]{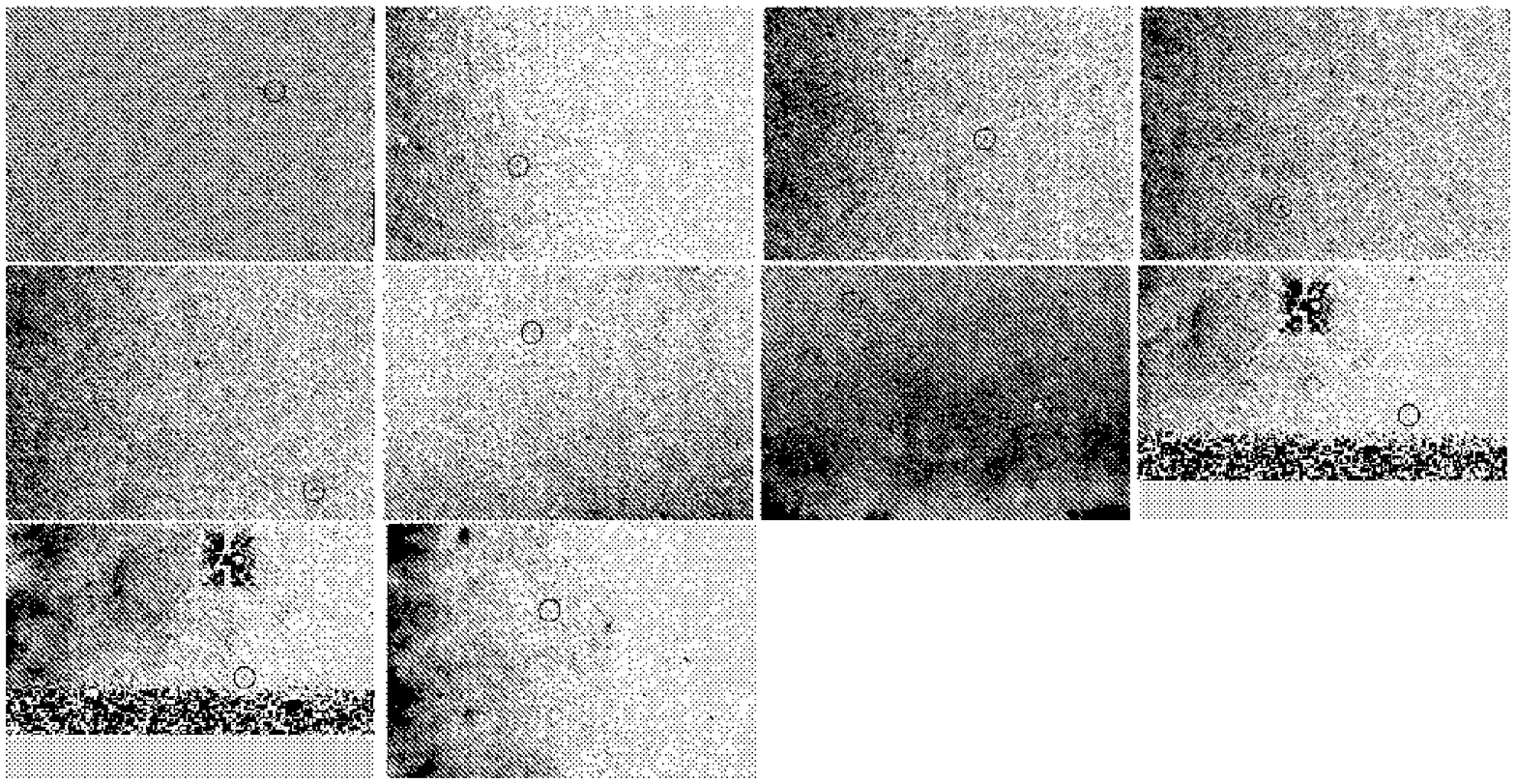}
\caption{The ten candidate halo PNe in our four WF/PC fields. The positions 
of the PNe are marked by black circles. From top left to bottom middle we 
show the fields of the PNe \#33, \#20, \#26, \#01, \#04, \#11, \#03, \#24, 
\#25 and \#13 of \citet{Ci91}.
\label{f:pn}}
\end{figure*}

Another source of H$\alpha$ emission in the halo are planetary nebulae (PNe). 
\citet{Ci91} detected in their ground-based study of NGC\,891 a total of 33 
PNe in the halo. Of these, 10 PNe are located in our WFPC2 field of view. 
We show our individual WFPC2 image sections with the candidate PNe marked 
by circles in Figure~\ref{f:pn}. 

Although the detections by \citet{Ci91} were based on the [\ion{O}{3}] 
emission line flux, which is usually the strongest emission line in PNe, 
those species often also exhibit a considerable amount of H$\alpha$ emission. 
We list the properties of the PNe in Table~\ref{t:halopn}, including 
identifier, position, size and distance from the midplane. The identifier 
is a combination of the WF/PC field and the running order sorted in R.A. for 
each of the four WF/PC fields. Of the 10 candidate PNe we marginally detected 
six PNe, with only two clear detections. From one of those two (PN-WF2-4) we 
estimated the flux in an aperture with a radius of $0\farcs5$ pixels, which 
yielded a magnitude of $\rm{m_{F656N}} = 26.13\pm0.01$\,mag, which 
corresponds roughly to the limiting magnitude of a point source we reached 
in our H$\alpha$ observation.

\begin{table*}
\begin{center}
\caption{Planetary nebulae in the halo of NGC\,891\label{t:halopn}}
\begin{tabular}{lcccccc}
\tableline
\tableline
Identifier & R.A.(J2000.0) & Dec.(J2000.0) & $|z|$ & size & 
detect. & alternate name\\
& [hh mm ss.ss] & [\degr\,\,\arcmin\,\,\arcsec] & [pc] &  & 
& \\ 
\tableline
PN-PC-1 & 02 22 38.96 & +42 22 25.5 & 1105 & ... & no & CJH91\#33\\
PN-WF2-1 & 02 22 29.67 & +42 22 21.9 & 1161 & unresolved & yes & 
CJH91\#20\\
PN-WF2-2 & 02 22 37.81 & +42 21 26.7 & 1520 & unresolved & yes & 
CJH91\#26\\
PN-WF2-3 & 02 22 37.90 & +42 21 48.7 & 1161 & unresolved & yes & 
CJH91\#01\\
PN-WF2-4 & 02 22 39.20 & +42 21 42.5 & 1934 & unresolved & yes & 
CJH91\#04\\
PN-WF3-1 & 02 22 31.80 & +42 21 46.7 & 1879 & ... & no & CJH91\#11\\
PN-WF3-2 & 02 22 33.71 & +42 22 03.3 & 1216 & ... & no & CJH91\#03\\
PN-WF4-1 & 02 22 33.08 & +42 22 20.6 & 1838 & unresolved & yes & 
CJH91\#24\\
PN-WF4-2 & 02 22 33.73 & +42 22 17.1 & 1442 & ... & no & CJH91\#25\\
PN-WF4-3 & 02 22 35.56 & +42 22 49.7 & 1133 & unresolved & yes & 
CJH91\#13\\
\tableline
\end{tabular}                   
\end{center}
\end{table*}

\section{COMPARISON WITH OTHER MULTIFREQUENCY HIGH SPATIAL RESOLUTION DATA}
\label{s:multif}

\subsection{Optical versus NIR narrowband imagery}
\label{ss:nirha}

The detailed investigation of the connection between the superbubbles and 
the filaments, that are driven by outflows, can be achieved by NIR narrowband 
observations, which trace the star formation such as the Br$\gamma$ or 
Pa$\alpha$ line. We have looked at the HST archival NICMOS Pa$\alpha$ 
observation of NGC\,891 for a detailed analysis of the disk-halo interface 
in NGC\,891. These observations were part of the wide field Pa$\alpha$ 
snapshot survey of nearby galaxy nuclear regions (P.I.: W.~Sparks), and have 
previously been published by \citet{Bo99}. The field of view of the NICMOS 
image is substantially ($\sim$ a factor of nine) smaller ($51\farcs2 \times 
51\farcs2$), compared to our WFPC2 image of NGC\,891. Unfortunaltely, the 
offsets of the image centers between the NICMOS and WFPC2 observations 
amount to $\Delta\alpha=+19\farcs5$ and $\Delta\delta=-81\farcs0$, and hence 
show only a very marginal overlap. 

The Pa$\alpha\,\lambda1.875\mu$m line observations trace the regions deep 
within the disk that are almost obscured at optical wavelengths due to the 
prominent dust lane. The image shows only a few brighter star-forming regions 
within the disk, and a moderately bright nuclear region. There is still a 
considerable amount of dust visible as well. As this pointing traces the 
nuclear part of NGC\,891, where already a substantial decrease in H$\alpha$ 
emission is visible from ground-based images, this confirms that there is 
much less star formation happening at and surrounding the nuclear regions, 
compared to the brighter emission regions in the northern part of NGC\,891 
\citep[e.g.,][]{De90,Ra90}.

\begin{figure*}[th]
\centering
\includegraphics[width=7in,angle=0,clip=t]{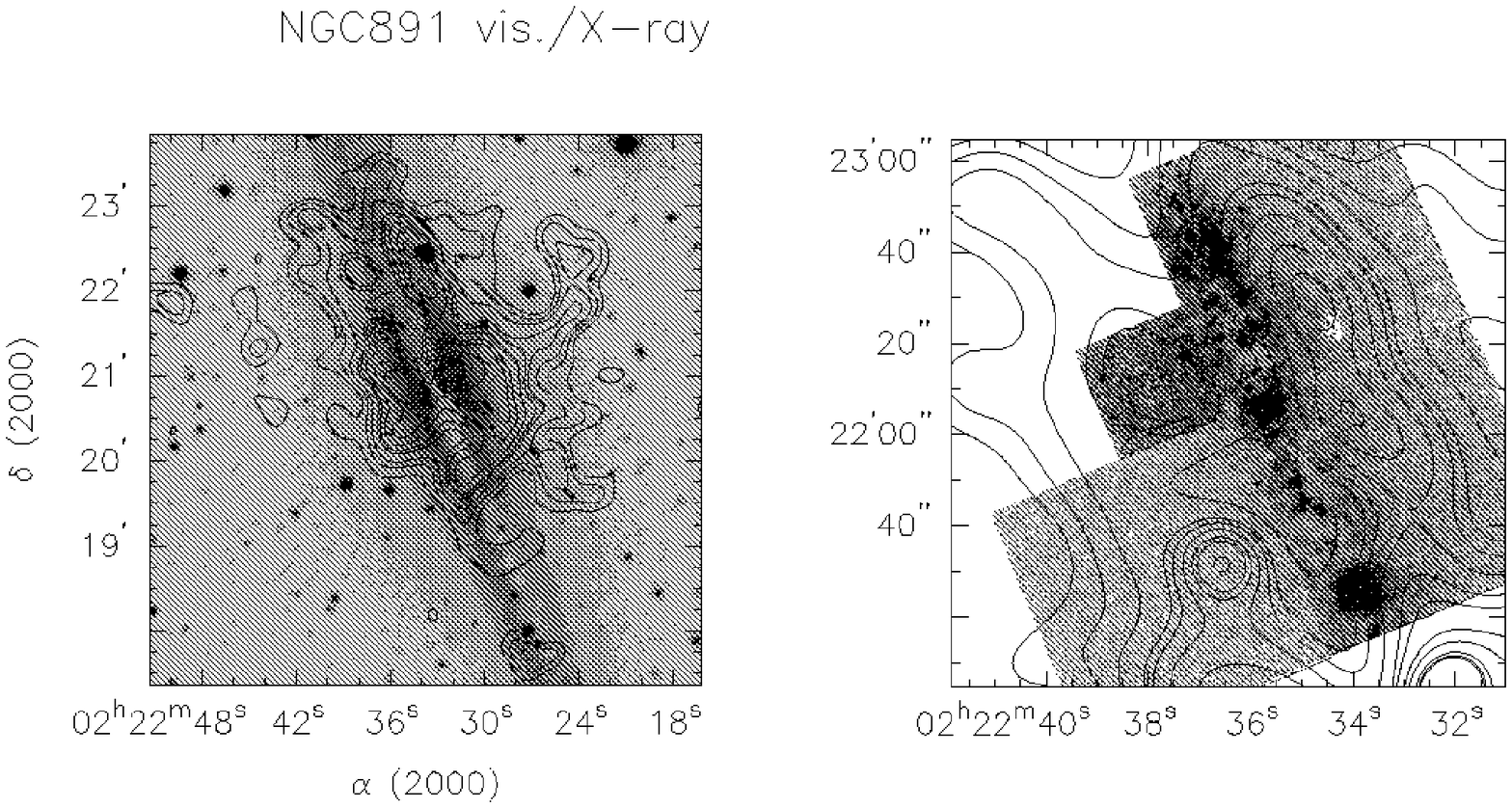}
\caption{The left panel shows the DSS image of NGC\,891, superimposed with 
the Chandra ACIS soft X-ray contours in the 0.3-2.0\,keV band. Contour levels 
are 1, 1.5, 2, 2.7, 3, 3.5, 4.1, 4.5, 5.5, 6.5, 8.5, 9.5, 10, 20 and 50 in 
units of 10$^{-9}$ photons\,$\rm{s^{-1}\,cm^{-2}\,arcsec^{-2}}$ (per pixel). 
The right panel shows a more detailed view with our WFPC2 H$\alpha$ image. 
\label{f:n891optxray}}
\end{figure*}

\subsection{A detailed comparison with Chandra X-ray data}
\label{ss:xrayha}

NGC\,891 has been studied in the X-ray regime with ROSAT \citep{Br94}, where 
an extended X-ray halo has been detected. More recently, within a sample of 
star-forming disk galaxies, the spatial and spectral properties of NGC\,891 
have been thoroughly investigated \citep{St04a,St04b}. We used the Chandra 
ACIS image of \citet{St04a} to compare the small scale structure of the hot 
ionized medium with the warm ionized medium. In Figure~\ref{f:n891optxray} 
we show the superposition of the two ISM constituents. The left panel shows 
the DSS image of NGC\,891 superimposed with the contours of the Chandra ACIS 
observation. The Chandra image shows the soft X-ray emission in the 
0.3-2.0\,keV band, which has been background subtracted and adaptively 
smoothed to achieve a better S/N ratio. The point sources have been removed 
as well. For further details on the X-ray observations and reductions we 
refer to \citet{St04a}. In the right hand panel we show the superposition of 
our WFPC2 H$\alpha$ image, again with the Chandra ACIS soft 0.3-2.0\,keV 
band. The global distribution of the X-ray emission follows the N-S asymmetry 
which was also observed in H$\alpha$ \citep[e.g.,][]{De90,RD02b} and in the 
radio continuum \citep[e.g.][]{Da94}. 

The disk of NGC\,891 lacks strong X-ray emission, although it is embedded in 
a pervasive soft X-ray layer. However, two local maxima in the 0.3-2.0\,keV 
band were detected in the northwestern halo region, above strong supershells 
in the disk. This is also the halo region ($|z|\geq$1\,kpc), where our 
diffuse H$\alpha$ emission is brightest. The pipeline processed XMM-Newton 
EPIC pn-image shows more or less the same global structure as the Chandra 
image. 

\section{DISCUSSION}
\label{s:discuss}

The general eDIG morphology of previous ground-based observations of the 
disk-halo interface in NGC\,891 is recovered \citep[e.g.,\,][]{De90,Ra90}. 
However, our high spatial resolution observation reveals an interesting 
sub-structure that has eluded ground-based observations. There is a 
pervasive eDIG layer detected. The observed H$\alpha$ morphology in NGC\,891 
is mostly diffuse, suggesting that a break-up of the homogeneous layer may 
only occur at very faint intensities with associated emission measures of 
less than 20\,$\rm{cm^{-6}\,pc}$. Several faint filaments are detected, 
which reach very high galactic latitudes $|z|\sim2$\,kpc. Some of these 
filaments are very collimated, and show no considerable broadening towards 
high-$|z|$. This may indicate, that magnetic fields have a strong influence 
on those features, and in turn can play a vital role for the transport 
mechanism. Since we do not see much evidence for the chimney scenario, 
other mechanisms have to be considered. As dust is also detected at 
high-$|z|$, though not spatially correlated with the eDIG on a one to one 
basis, soft mechanisms for the transport of the dust are favored such as {\em 
photolevitation} \citep{Fr91,Fe91}, or possibly magnetic field instabilities 
\citep{Pa92,Sh01}. However, which of the processes act as the actual 
transport of the dust features is not known yet. Whereas the eDIG morphology 
is smoothly distributed, the dusty ISM phase shows much more sub-structure 
and is generally very clumpy. In fact, a very intricate network of dust 
features is observed which shows an irregular pattern. The detected ionized 
structures (i.e. the supergiant shells) in the galactic midplane - in 
contrast to the high-$|z|$ dusty filaments - are comparable in size and 
shape to the Milky Way analogons \citep[e.g.,\,][]{Hei84,Ko92}.

One of the key results of our study is that the DIG does not break up into 
many chimneys at the heights above the disk where we can see it. More 
specifically, this is true for our spatial resolution and sensitivity limits 
(5\,pc width and a surface brightness above 20\,cm$^{-6}$\,pc). This can 
have various reasons. It is possible that the medium is intrinsically 
too unorganized (i.e. inhomogeneous) for organized structures like chimneys 
to survive or to be formed in the first place. That is, the outflow of hot 
gas occurs mostly through an irregular medium, maybe fractal in nature 
\citep[e.g.,][]{El98}, such that the energy can get out quickly enough that 
the expected formation of walls is not happening at all. It has been argued 
that only in starbursts sufficient energy is produced locally that matter 
may be piled up in chimneys \citep{Wi99}. 

The lack of correlation of dust and H$\alpha$ filaments may also point to a 
much more chaotic medium than could be modeled in the chimney scenario. On 
the other hand, the dust appears more filamentary than the H$\alpha$. This 
could be due to the difference in morphology. We are likely seeing
preferentially denser dust features and not the diffuse dust. 

Given the sensitivity limit of EM = 20\,cm$^{-6}$\,pc, we can roughly 
estimate to what kind of chimney wall, seen edge-on, this would correspond 
to. Our spatial resolution limit is 5\,pc. For an assumed 20\,pc long chimney 
wall, one needs a medium with a density of $n_{\rm{e}}$ = 1\,cm$^{-3}$. This 
is not regarded as a very dense medium. If the density is to be much higher, 
the chimney wall would have to be much thinner or the pathlength along it 
has to be much less than 20\,pc. Alternatively, the chimneys might indeed be 
there, {\em but only} closer to the disk, where we cannot see them, possibly 
due to heavy dust obscuration. 

The outflow, as in a starburst galaxy, is dominated by a few major complexes,
which we indeed also see below the few longer filaments in NGC\,891. These 
could be chimney walls or remnants of chimney walls. Alternatively, the 
larger filaments could indicate some large scale magnetic field topology, in 
which the magnetic field would be responsible for the organization of matter 
in them, not necessarily the chimneys. In order to vindicate or refute this 
conjecture, much higher spatial resolution and higher sensitivity radio 
continuum observations are mandatory. 

\subsection{Collimated filaments - magnetically confined?}
\label{ss:images}

Quite surprisingly, there are these few very narrow filaments discerned, 
which are well collimated, even at high galactic latitudes. The chimney 
structures, or at least the walls of chimneys, which would be observationally 
visible as dust pillars, would show a widening with increasing distance from 
the galactic midplane. However, we do not see such structures in our 
H$\alpha$ image. Instead, these very collimated long filaments seem to 
originate from the bright superbubbles in the galactic midplane. In a Milky 
Way study \citet{Haf98} found two very collimated long filaments based on 
sensitive observations carried out with the Wisconsin H$\alpha$ Mapper 
(WHAM). One of the two faint filaments they discovered is 80\degr\, long 
(corresponding to a length of about 1.2\,kpc above the midplane). It has a 
nearly constant H$\alpha$ intensity profile of EM$\approx1.1\,\rm{cm^{-6}\,
pc}$. Although the origin of the filaments is not quite clear, a jetlike 
ejection from the CMa R1/OB1 association has been speculated, but other 
mechanisms seem possible too \citep{Haf98}. So it may well be that the 
filaments seen in the Milky Way and those identified by us in NGC\,891 have 
a similar, yet unknown origin.   

In a study of NGC\,891 by \citet{Sc96} they used optical polarization 
measurements and they found vertical magnetic fields in the halo region 
above the inner part (i.e. the bulge region). They speculated about an 
optical spur like the North Galactic Spur known in the Milky Way 
\citep{Ma70}. However, optical polarization measurements faces challenges, 
as it might be difficult, if possible at all, to disentangle whether the 
polarization is due to grain alignment or by dust scattering. Quite 
interestingly, the maximum polarization vector (perpendicular to the disk) 
in these observations \citep{Sc96} is close to the position above our 
detected supergiant shell, from which the collimated features rise into the 
halo. 

\section{SUMMARY}
\label{s:sum}

We have presented new high spatial resolution narrowband observations of 
eDIG in the nearby edge-on galaxy NGC\,891, obtained with the WFPC2 
on-board HST. A region 6.9\,kpc$\times6.9$\,kpc in size, centered on the 
northeastern part of NGC\,891, was investigated. An extraplanar layer with 
an extension of at least 1.8\,kpc above/below the galactic midplane was 
found. Superimposed onto this eDIG layer are several filaments and arcs, 
reaching extraplanar distances of about $|z|\approx2.2$\,kpc. A few 
superthin, and collimated filaments were detected which have a surprisingly 
high length/width ratio of up to $l/b \sim140$! Most part of the H$\alpha$ 
emission is diffuse, however, several discrete emission features are 
detected. Whereas the observed global morphology is validating the results 
of previous ground-based observations, the high spatial resolution 
observations reveal local structures that have not been resolved or detected 
by ground-based optical observations. The mechanisms of the gas and energy 
transport into the halo cannot be verified without ambiguity, however, the 
chimney scenario is most likely not responsible for the transport, as no 
clear structures representing individual chimneys (i.e. the walls of 
chimneys) have been detected. The highly collimated gaseous filaments suggest 
an influence by magnetic fields. We have furthermore presented results of 
yet another constituent of the ISM, in the disk-halo interface of NGC\,891, 
the dust. 

Extraplanar dust was detected, and has been analyzed from the unsharp-masked 
broadband image. The observations reveal extraplanar dust at high galactic 
latitudes, as previously reported by \citep{HoSa97,HoSa00}. We detected 
dusty filaments, up to $|z|\sim2$\,kpc above the galactic midplane. Although 
there is a correlation observed between the {\it two} different ISM 
constituents, as both constituents are detected in the halo of NGC\,891, 
however, generally no one to one correlation between individual filaments and 
dust filaments are observed. The dust morphology is twofold. Many dust 
filaments are aligned perpendicular to the galactic disk, indicating that 
magnetic fields may play a key role in transporting the dust into the 
disk-halo interface. Other dusty filaments, however, located mostly at the 
highest galactic latitudes, are aligned parallel to the disk, which may 
represent a temporal equilibrium of the dusty ISM phase, possibly transported 
by soft mechanisms (e.g.,\,photolevitation) into the galactic halo. 

The spatial distribution of the soft X-ray emission in the halo of NGC\,891, 
as revealed by Chandra, is in good agreement with our H$\alpha$ observations. 
There are two local maxima of X-ray emission in the halo, where also the 
diffuse H$\alpha$ emission is strongest.

\acknowledgments

We would like to thank David Strickland for providing us with the calibrated 
Chandra ACIS image of NGC\,891. We also thank the anonymous referee for a 
swift report and helpful comments which improved the clarity of the paper in 
a few places. The authors (JR and RJD) acknowledge financial support for this 
research project by the Deutsches Zentrum f\"ur Luft- und Raumfahrt (DLR) 
through grant 50\,OR\,9707.


\begin{thebibliography}{}

\bibitem[Alton et al.(1998)]{Alton98} 
Alton, P. B., Bianchi, S., Rand, R. J., Xilouris, E. M., Davies, J. I., 
\& Trewhella, M. 1998 \apj, 507, L125

\bibitem[Birk et al.(1998)]{Bir98} 
Birk, G.T., Lesch, H., \& Neukirch, T. 1998, \mnras, 296, 165

\bibitem[B\"oker et al.(1999)]{Bo99}
B\"oker, T., Calzetti, D., Sparks, W., Axon, D., Bergeron, L. E., Bushouse, 
H., et al. 1999, \apjs, 124, 95

\bibitem[Bregman \& Pildis(1994)]{Br94} 
Bregman, J. H., \& Pildis, R. A. 1994, \apj, 420, 570

\bibitem[Cecil et al.(2001)]{Ce01} 
Cecil, G., Bland-Hawthorn, J., Veilleux, S., \& Filippenko, A. V. 2001, 
\apj, 555, 338

\bibitem[Cecil et al.(2002)]{Ce02} 
Cecil, G., Bland-Hawthorn, J., \& Veilleux, S. 2002, \apj, 576, 745

\bibitem[Ciardullo et al.(1991)]{Ci91} 
Ciardullo, R., Jacoby, G. H., \& Harris, W. H. 1991, \apj, 383, 487

\bibitem[Chevalier \& Clegg(1985)]{Ch85} 
Chevalier, R. A., \& Clegg, A. W. 1985, \nat, 317, 44

\bibitem[Collins et al.(2000)]{Co00} 
Collins, J. A., Rand, R. J., Duric, N., \& Walterbos, R. A. M. 2000, \apj, 
536, 645

\bibitem[Dahlem et al.(1994)]{Da94} 
Dahlem, M., Dettmar, R.-J., \& Hummel, E. 1994, \aap, 290, 384

\bibitem[de Avillez(2000)]{Avi00} 
de Avillez, M. A. 2000, \mnras, 315, 479  

\bibitem[Dettmar(1990)]{De90} 
Dettmar, R.-J. 1990, \aap, 232, L15

\bibitem[Dettmar \& Schulz(1992)]{De92} 
Dettmar, R.-J., \& Schulz, H. 1992, \aap, 254, L25

\bibitem[Domg\"orgen \& Mathis(1994)]{DoMa94} 
Domg\"orgen, H., \&  Mathis, J. S. 1994, \apj, 428, 647

\bibitem[Dove \& Shull(1994)]{DoSh} 
Dove, J. B., \& Shull, J. M. 1994, \apj, 430, 222

\bibitem[Elmegreen(1998)]{El98}
Elmegreen, B. G. 1998, \pasa, 15, 74

\bibitem[Ferguson et al.(1996a)]{Fer96}
Ferguson, A. M. N., Wyse, R. F. G., \& Gallagher, J. S. 1996a, \aj, 112, 2567

\bibitem[Ferguson et al.(1996b)]{Fer96b}
Ferguson, A. M. N., Wyse, R. F. G., Gallagher, J. S., \& Hunter, D. A. 
1996b, \aj, 111, 2265

\bibitem[Ferrara et al.(1996)]{Fe96} 
Ferrara, A., Bianchi, S., Dettmar, R.-J., \& Giovanardi, C. 1996, \apj, 
467, L69

\bibitem[Ferrara et al.(1991)]{Fe91} 
Ferrara, A., Ferrini, F., Barsella, B., \& Franco, J. 1991, \apj, 381, 137

\bibitem[Franco et al.(1991)]{Fr91} 
Franco, J., Ferrini, F., Barsella, B., \& Ferrara, A. 1991, \apj, 366, 443

\bibitem[Fruchter \& Hook(2002)]{FrHo02} 
Fruchter, A. S., \& Hook, R. N. 2002, \pasp, 114, 144

\bibitem[Garc\'{\i}a-Burillo et al.(1992)]{Ga92} 
Garc\'{\i}a-Burillo, S., Gu{\'e}lin, M., Cernicharo, J., \&  Dahlem, M. 
1992, \aap, 266, 21

\bibitem[Gaustad et al.(2001)]{Ga01}
Gaustad, J. E., McCullough, P. R., Rosing, W., \& Van Buren, D. 2001, \pasp, 
113, 1326

\bibitem[Gerhard et al.(2002)]{Ge02}
Gerhard, O., Arnaboldi M., Freeman, K. C., \& Okamura, S. 2002, \apj, 580, 
L121

\bibitem[Gonzaga et al.(1998)]{Go98} 
Gonzaga, S., Biretta, J., Wiggs, J. C., Hsu, J. C., Smith, T. E., Bergeron, 
L., \& the STScI WFPC2 group, 1998, The Drizzling Cookbook, Instrument 
Science Report WFPC2 98-04, STScI

\bibitem[Haffner et al.(1998)]{Haf98}
Haffner, L. M., Reynolds, R. J., \& Tufte, S. L. 1998, \apj, 501, L83

\bibitem[Haffner et al.(2003)]{Haf03}
Haffner, L. M., Reynolds, R. J., Tufte, S. L., Madsen, G. J., Jaehnig, K. P., 
\& Percival, J. W. 2003, \apjs, 149, 405

\bibitem[Hambly et al.(1996)]{Ham96}
Hambly, N. C., Wood, K. D., Keenan, F. P., Kilkenny, D., Dufton, P. L., 
\& Miller, L., et al. 1996, \aap, 306, 119

\bibitem[Heiles(1984)]{Hei84} 
Heiles, C. 1984, \apjs, 55, 585

\bibitem[Hoopes et al.(1999)]{Ho99} 
Hoopes, C. G., Walterbos, R. A. M., \& Rand, R. J. 1999, \apj, 522, 669 

\bibitem[Howk \& Savage(1997)]{HoSa97} 
Howk, J. C., \& Savage, B. D. 1997, \aj, 114, 2463

\bibitem[Howk \& Savage(1999)]{HoSa99} 
Howk, J. C., \& Savage, B. D. 1999, \aj, 117, 2077

\bibitem[Howk \& Savage(2000)]{HoSa00} 
Howk, J. C., \& Savage, B. D. 2000, \aj, 119, 644

\bibitem[Kennicutt(1988)]{Ke88}
Kennicutt, R. C., Jr. 1988, \apj, 334, 144

\bibitem[Keppel et al.(1991)]{Ke91} 
Keppel, J. W., Dettmar, R.-J., Gallagher, J. S., III, \& Roberts, M. S. 
1991, \apj, 374, 507

\bibitem[Kilkenny et al.(1995)]{Ki95}
Kilkenny, D., Luvhimbi, E., O'Donoghue, D., Stobie, R. S., Koen, C., \& 
Chen, A. 1995, \mnras, 276, 906

\bibitem[Koo et al.(1992)]{Ko92} 
Koo, B.-C., Heiles, C., \& Reach, W. T. 1992, \apj, 390, 108 

\bibitem[Korpi et al.(1999)]{Ko99} 
Korpi, M. J., Brandenburg, A., Shukurov, A., Tuominen, I., \& Nordlund, \AA. 
1999, \apj, 514, L99

\bibitem[Lehnert \& Heckman(1995)]{LeHe95} 
Lehnert, M. D., \& Heckman, T. M. 1995, \apjs, 97, 89

\bibitem[Mathewson \& Ford(1970)]{Ma70}
Mathewson, D. S., \& Ford, V. L. 1970, \memras, 74, 139

\bibitem[Mathis(1986)]{Ma86} 
Mathis, J. S. 1986, \apj, 301, 424

\bibitem[Meaburn(1980)]{Mea80}
Meaburn, J. 1980, \mnras, 192, 365

\bibitem[Miller \& Veilleux(2003)]{MiVe03} 
Miller, S. T., \& Veilleux, S. 2003, \apjs, 148, 383

\bibitem[Miller \& Cox(1993)]{MiCo93} 
Miller, W. W., III, \& Cox, D. P. 1993, \apj, 417, 579 

\bibitem[Norman \& Ikeuchi(1989)]{NoIk} 
Norman, C. A., \& Ikeuchi, S. 1989, \apj, 345, 372

\bibitem[O'Dell \& Doi(1999)]{Od99}
O'Dell, C. R., \& Doi, T. 1999, \pasp, 111, 1316

\bibitem[Parker(1992)]{Pa92} 
Parker, E. N. 1992, \apj, 401, 137

\bibitem[Pildis et al.(1994)]{Pi94} 
Pildis, R. A., Bregman, J. N., \& Schombert, J. M. 1994, \apj, 427, 160 

\bibitem[Rand(1996)]{Ra96} 
Rand, R. J. 1996, \apj, 462, 712

\bibitem[Rand(1997)]{Ra97} 
Rand, R. J. 1997, \apj, 474, 129

\bibitem[Rand(1998)]{Ra98} 
Rand, R. J. 1998, \apj, 501, 137

\bibitem[Rand(2000)]{Ra00} 
Rand, R. J. 2000, \apj, 537, L13

\bibitem[Rand et al.(1990)]{Ra90} 
Rand, R. J., Kulkarni, S. R., \& Hester, J. J. 1990, \apj, 352, L1

\bibitem[Reynolds(1984)]{Rey84} 
Reynolds, R. J. 1984, \apj, 282, 191

\bibitem[Reynolds et al.(2001)]{Rey01} 
Reynolds, R. J., Sterling, N. C., Haffner, L. M., \& Tufte, S. L. 
2001, \apj, 548, L221

\bibitem[Rossa \& Dettmar(2000)]{RD2000} 
Rossa, J., \& Dettmar, R.-J. 2000, \aap, 359, 433

\bibitem[Rossa \& Dettmar(2003a)]{RD02a} 
Rossa, J., \& Dettmar, R.-J. 2003a, \aap, 406, 493 

\bibitem[Rossa \& Dettmar(2003b)]{RD02b} 
Rossa, J., \& Dettmar, R.-J. 2003b, \aap, 406, 505 

\bibitem[Rupen(1991)]{Ru91} 
Rupen, M. P. 1991, \aj, 102, 48

\bibitem[Ryan-Weber et al.(2004)]{Rya04}
Ryan-Weber, E. V., Meurer, G. R., Freeman, K. C., Putman M. E., Webster, 
R. L., \& the SINGG team 2004, AJ, 127, 1431 

\bibitem[Scarrott \& Draper(1996)]{Sc96}
Scarrott, S. M., \& Draper, P. W. 1996, \mnras, 278, 519

\bibitem[Shapiro \& Field(1976)]{Sha76} 
Shapiro, P. R., \& Field, G. B. 1976, \apj, 205, 762

\bibitem[Shchekinov et al.(2001)]{Sh01} 
Shchekinov, Yu. A., Dettmar, R.-J., Schr\"oer, A., \& Steinacker, A., 
2001, Astron. and Astrophys. Transact., 20, 237 (see also astro-ph/0102164)

\bibitem[Slavin et al.(1993)]{Sl93} 
Slavin, J. D., Shull, J. M., \& Begelman, M. C. 1993, \apj, 407, 83 

\bibitem[Strickland et al.(2004a)]{St04a}
Strickland, D. K., Heckman, T. M., Colbert, E. J. M., Hoopes, C. G., \&  
Weaver, K. A. 2004a, \apjs, 151, 193

\bibitem[Strickland et al.(2004b)]{St04b}
Strickland, D. K., Heckman, T. M., Colbert, E. J. M., Hoopes, C. G., \&  
Weaver, K. A. 2004b, \apj, 606, 829 

\bibitem[Swaters et al.(1997)]{Sw97} 
Swaters, R. A., Sancisi, R., \&  van der Hulst, J. M. 1997, \apj, 491, 140

\bibitem[Thilker et al.(2002)]{Th02}
Thilker, D. A., Walterbos, R. A. M., Braun, R., \& Hoopes, C. G. 2002, 
\aj, 124, 3118

\bibitem[T\"ullmann et al.(2000)]{Tu00} 
T\"ullmann, R., Dettmar, R.-J., Soida, M., Urbanik, M., \& Rossa, J. 
2000, \aap, 364, L36

\bibitem[T\"ullmann et al.(2003)]{Tu03} 
T\"ullmann, R., Rosa, M. R., Elwert, T., Bomans, D. J., Ferguson, A. M. N., 
\& Dettmar, R.-J. 2003, \aap, 412, 69

\bibitem[van den Bergh(1992)]{vdB92} 
van den Bergh, S. 1992, \pasp, 104, 861

\bibitem[Wada(2001)]{Wa01} 
Wada, K. 2001, \apj, 559, L41

\bibitem[Wada \& Norman(1999)]{WaNo99} 
Wada, K., \& Norman, C. A. 1999, \apj, 516, L13

\bibitem[Wada \& Norman(2001)]{WaNo01} 
Wada, K., \& Norman, C. A. 2001, \apj, 547, 172

\bibitem[Walterbos \& Braun(1994)]{Wal94} 
Walterbos, R. A. M., \& Braun R. 1994, \apj, 431, 156  

\bibitem[Wang et al.(2001)]{Wan01} 
Wang, Q. D., Immler, S., Walterbos, R. A. M., Lauroesch, J. T., \& 
Breitschwerdt, D. 2001, \apj, 555, L99 

\bibitem[Wills et al.(1999)]{Wi99}
Wills, K. A., Redman, M. P., Muxlow, T. W. B., \& Pedlar, A. 1999, \mnras, 
309, 395

\end{thebibliography}
\end{document}